\documentclass[aps,english,twocolumn]{revtex4-1}
\usepackage[T1]{fontenc}
\usepackage[latin9]{inputenc}
\usepackage{graphicx}
\usepackage{esint}
\usepackage{color}
\usepackage{lipsum}
\usepackage{subfigure}
\usepackage{mwe}
\makeatletter

\usepackage{babel}

\makeatother

\usepackage{babel}
\begin{document}
\title{Dissipative soliton interaction in Kerr resonators with high-order dispersion }
\author{A. G. Vladimirov$^{a}$, M. Tlidi$^{b}$, and M. Taki$^{c}$}
\affiliation{$^{a}$Weierstrass Institute, Mohrenstrasse 39, 10117 Berlin, Germany}
\affiliation{$^{b}$Département de Physique, Faculté des Sciences, Université Libre
de Bruxelles (U.L.B.), CP 231, Campus Plaine, B-1050 Bruxelles, Belgium}
\affiliation{$^{c}$Université de Lille, CNRS, UMR 8523 - PhLAM - Physique des
Lasers Atomes et Molécules, F-59000 Lille, France}
\begin{abstract}
We consider an optical resonator containing a photonic crystal fiber and driven coherently by an injected beam. This device is described by a generalized Lugiato-Lefever equation with fourth order dispersion. 
We use an asymptotic approach to derive interaction equations  governing the slow time evolution  of the coordinates of two interacting dissipative solitons. We show that Cherenkov radiation induced by positive fourth-order dispersion leads to a strong increase of the interaction force between the solitons. As a consequence,  large number of equidistant soliton bound states in the phase space of the interaction equations can be stabilized.  We show that the presence of even small spectral filtering not only dampens the Cherenkov radiation at the soliton tails and  reduces the interaction strength, but can also affect the the bound state stability.  
\end{abstract}
\maketitle
{\section{Introduction}}.
Optical frequency combs generated by micro-cavity resonators have revolutionized many fields of science and technology,  such as high-precision
spectroscopy, metrology, and photonic analog-to-digital conversion
\citep{fortier201920}. A particular interest is paid to the soliton
frequency combs associated with the formation in the time domain of
the so-called temporal cavity solitons -- nonlinear localized structures
of light, which preserve their shape in the course of propagation.
Temporal dissipative solitons often called cavity solitons were reported experimentally in mode-locked lasers, micro-cavity resonators \citep{kippenberg2018dissipative,Herr14},
and in coherently driven fiber cavities \citep{Leo10}. 

In this work we consider a photonic crystal fiber cavity driven by a coherent injected beam. When operating close to the zero dispersion wavelength, high-order chromatic dispersion effects could play an important role in the dynamics of the system. Taking into account these effects together with spectral filtering the dimensionless model equation in the mean-field limit reads 
\begin{equation}
\partial_{t}U=S-(1+i\theta)U+iU|U|^{2}+\left(\delta+i\right)\partial^{2}_{\tau}U+\beta_{3}\partial^{3}_{\tau}U+i\beta_{4} \partial^{4}_{\tau}U,\label{eq:LL}
\end{equation}
where $U(\tau,t)$ is the complex electric field envelope, $\tau$
is time and $t$ is the slow time variable describing the number of round trips in the cavity. The parameter $S$ measures the injection rate, $\theta$ describes frequency detuning, second order dispersion and Kerr nonlinearity coefficients are normalized to unity, $\beta_{3}$
and $\beta_{4}$ are the third and fourth-order dispersion coefficients, respectively, and $0<\delta\ll1$ is the small spectral filtering coefficient. The optical losses are determined by the mirror transmission and the intrinsic material absorption. This losses are normalized to unity.

In the absence of high-order dispersion and spectral filtering, we recover from Eq. (\ref{eq:LL}) the Lugiato-Lefever equation \citep{LL87} which is a paradigmatic model to study temporal cavity solitons (see overview  \citep{Chembo2017theory,lugiato2018lugiato}). It is widely applied to describe two important physical systems: passive ring fiber cavity with coherent optical injection and driven optical microcavity used for frequency comb generation \citep{haelterman1992dissipative,maleki2010high,matsko2011mode,Chembo2013}. 
The inclusion of the fourth-order dispersion allows the modulational instability to have a finite domain of existence delimited by two pump power values \citep{tlidi2007control}. As a consequence, upper homogeneous steady state solution becomes modulationally stable and dark dissipative solitons sitting this  solution can appear \citep{tlidi2010high}. In the presence of third order dispersion bright and dark dissipative solitons become asymmetric and acquire an additional group velocity shift associated with this asymmetry \citep{akhmediev1990modulation,tlidi2013drift,Milian2014,vladimirov2018effect}.

Being well separated from one another dissipative solitons can interact via their exponentially decaying tails and form bound states characterized by fixed distances
between the solitons. This weak interaction can be strongly affected by different perturbations, such as  periodic modulation \citep{Soto03,TVZ12} and high-order dispersions \citep{Oliver06}, which lead to the appearance of the so-called soliton Cherenkov radiation at the soliton tails \citep{Akhmediev95}. Soliton interaction in the presence of high-order dispersions was theoretically studied  in several works  \citep{Oliver06,Milian2014,parra2016dark,parra2017interaction,vladimirov2018effect}.
However, they were either focused on the asymmetric soliton interaction in the presence of third-order dispersion or based mainly on the numerical calculation of the soliton interaction potential. Unlike these works, here we present an analytical theory of the interaction of two dissipative solitons
of the Lugiato-Lefever equation with fourth-order dispersion term based on the asymptotic approach developed in \citep{gorshkov1981interactions,karpman1981perturbational}.
Furthermore, we show that similarly to the case of the interacting
oscillatory solitons \citep{TVZ12}, a small spectral filtering effect can strongly affect the interaction force and the stability properties of the bound soliton states.

\section{Single peak dissipative soliton}
Without high-order dispersion and spectral filtering terms, $\beta_{3}=\beta_{4}=\delta=0$, Eq. (\ref{eq:LL}) supports a single or multipeak dissipative solitons characterized by damped oscillatory tails   \citep{scroggie1994pattern}.  Stable dissipative solitons have been found in a strongly nonlinear regime, where the modulational instability is subcritical, i.e.,  for $\theta>41/30$. More precisely, they have been found in the pinning region, where the lower stationary homogeneous solution coexist with a periodic one. The number of dissipative solitons and their distribution in the cavity are determined by the initial conditions while their maximum peak power remains constant for fixed values of the system parameters  \citep{scroggie1994pattern}. For $\theta>41/30$ Eq. (\ref{eq:LL}) supports a single peak dissipative soliton solution in the
form $U(t,\tau)=U_{0}+u_{0}(\tau)$, where $I_{0}=|U_{0}|^{2}=const$ is the intensity of the stationary homogeneous solution of Eq. (\ref{eq:LL}) and $u_{0}(\tau)$ decays exponentially at $\tau\to\pm\infty$. This solution persists also at sufficiently small $\beta_{3}$, $\beta_{4}$, and $\delta$. It remains motionless for $\beta_{3}=0$ and becomes uniformly moving otherwise, $U(t,\tau)=U_{0}+u_{0}(\tau-vt)$.
Asymptotic analytic theory of the asymmetric dissipative soliton interaction via Cherenkov radiation induced by the third-order dispersion coefficient $\beta_{3}$ was developed in \citep{vladimirov2018effect}.
Below we consider the case when only small fourth-order dispersion is present, $\beta_{3}=0$ and $|\beta_{4}|\ll1$. In this case due to the symmetry property of Eq.
(\ref{eq:LL}), $\tau\to-\tau$, the soliton velocity is always zero, $v=0$. 

\begin{figure}
\includegraphics[scale=0.33]{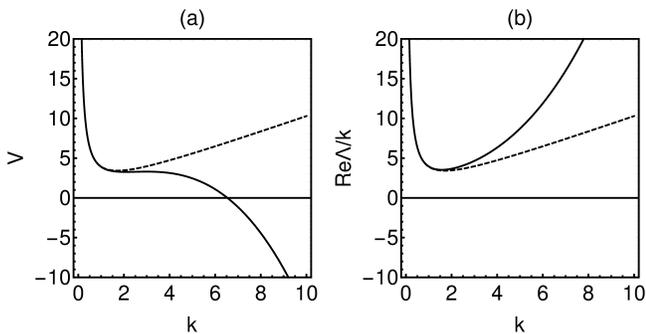}
\caption{Phase velocity $V$ of small dispersive waves with positive (a) and negative (b) fourth-order dispersion coefficient $\beta_{4}$, and $\beta_{3}=0$.  Solid line corresponds to $\beta_{4}=0.025$ (a) and $\beta_{4}=-0.025$ (b). Dashed line corresponds to $\beta_{4}=0$. The parameter values are $S=1.8$, $\theta=3.5$, and $\delta=0.02$.} 
\label{Fig0}
\end{figure}

The dispersion relation for the small  amplitude waves is determined by substituting $U(t,\tau)=U_{0}+A_{0}e^{ik\tau-i\Lambda t}$ into
Eq. (\ref{eq:LL}) and linearizing the resulting equation at $U=U_{0}$.
This yields 
\[
\Lambda=-2I_{0}^{2}+i\sqrt{\left(1+\delta k^{2}\right)^{2}-I_{0}^{2}}+k^{2}-\beta_{4} k^{4}.
\]
The phase velocity of the dispersive waves $V=\Re(\Lambda)/k$ is shown in Fig. \ref{Fig0}(a) for positive (\ref{Fig0}(a)) and negative (\ref{Fig0}(b)) $\beta_{4}$, as a function of the wave number $k$. Cherenkov radiation appears when the phase velocity $V$ coincides with zero soliton velocity as shown in Fig.\ref{Fig0}(a). It is seen from this figure that the Cherenkov radiation emitted from the soliton tail occurs only when $\beta_{4}$ is positive. Therefore, below we consider only the case of positive fourth-order dispersion coefficient $0<\beta_{4}\ll1$ when the Cherenkov radiation is present. For negative $\beta_{4}$ the soliton interaction is only weakly affected by the small fourth-order dispersion term.

\begin{figure}
\includegraphics[scale=0.35]{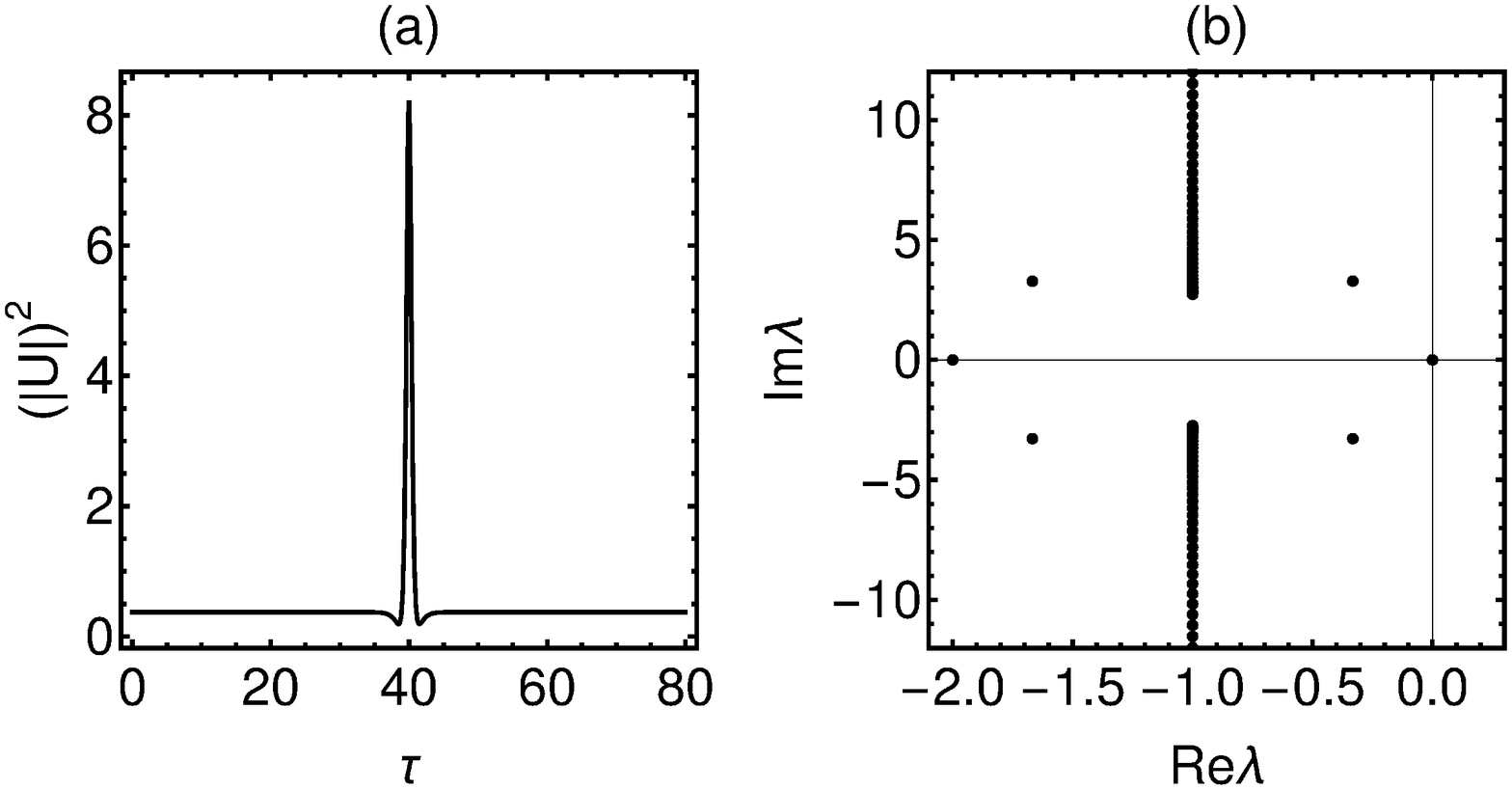}
\caption{Soliton solution of the Lugiato-Lefever equation (\ref{eq:LL}) with
$\beta_{3}=\beta_{4}=\delta=0$ (a) and eigenvalue spectrum obtained by
numerical linear stability analysis of this solution (b). Other parameters
are the same as in Fig. \ref{Fig0}.}
\label{Fig:soliton0}
\end{figure}

Linear stability of the dissipative soliton solution $u_{0}\left(\tau\right)$
is  determined by calculating the eigenvalue spectrum $\lambda$ of the operator
\begin{equation}
\hat{L}\left(\mathbf{u}_{0}\right)=\hat{L}_{0}+\hat{L}_{1}\left(\mathbf{u}_{0}\right),\label{eq:L}
\end{equation}
obtained by linearization of Eq. (\ref{eq:LL}) around the soliton solution.
Here $\mathbf{u}_{0}=\left(\begin{array}{c}
u_{0}\\
u_{0}^{*}
\end{array}\right)$ , $\hat{L}_{0}=\hat{L}(0)$ is the linear differential operator evaluated at the stationary homogeneous solution $\mathbf{U}=\mathbf{U}_{0}$:
\begin{widetext}
\[
\hat{L}_{0}=\left(\begin{array}{cc}
-1-i\theta+2iI_{0}^{2}+\left(i+\delta\right)\partial^{2}_{\tau}+i\beta_{4}\partial^{4}_{\tau} & iU_{0}^{2}\\
-iU_{0}^{*2} & -1+i\theta-2iI_{0}^{2}-\left(i-\delta\right)\partial^{4}_{\tau}-i\beta_{4}\partial^{4}_{\tau}
\end{array}\right),
\]
and 
\[
\hat{L}_{1}\left(\mathbf{u}_{0}\right)=\left(\begin{array}{cc}
2iU_{0}^{*}u_{0}+2iU_{0}u_{0}^{*}+2i|u_{0}|^{2} & 2iU_{0}u_{0}+iu_{0}^{2}\\
-2iU_{0}^{*}u_{0}^{*}-iu_{0}^{*2} & -2iU_{0}^{*}u_{0}+2iU_{0}u_{0}^{*}-2i|u_{0}|^{2}
\end{array}\right).
\]
\end{widetext}
We have calculated numerically the soliton solution and the eigenvalue spectrum $\lambda$ of the operator $\hat{L}\left(\mathbf{u}_{0}\right)$ by discretizing Eq. (\ref{eq:LL}) on an uniform grid of $2000$ points on the interval $\tau\in[0,80]$ with periodic boundary conditions. The result is shown in Fig. \ref{Fig:soliton0} for $\beta_{3}=\beta_{4}=\delta=0$. The continuous spectrum lies on the line $\Re(\lambda)=-1$, while the discrete spectrum of the soliton is symmetric with respect to this line \citep{barashenkov1996existence}. For the parameter values of Fig. \ref{Fig:soliton0} apart from two real eigenvalues: zero eigenvalue, $\lambda=0$, associated with the translational symmetry of the Lugiato-Lefever equation and symmetric one, $\lambda=-2$, soliton has two symmetric pairs of complex conjugated eigenvalues. The right pair of these complex eigenvalues is responsible for an Andronov-Hopf bifurcation taking place with the increase of the injection parameter
$S$. 
The decay rates of the soliton tails depend on the eigenvalues $\mu$
satisfying the characteristic equation 
\begin{widetext}
\begin{equation}
\beta_{4}^{2}\mu^{8}+2\beta_{4}\mu^{6}+\left[1+\delta^{2}+2\beta_{4}(2I_{0}-\theta)\right]\mu^{4}+2(2I-\theta-\delta)\mu^{2}+\left[1-I_{0}^{2}+(2I_{0}-\theta)^{2}\right]=0.\label{eq:characteristic}
\end{equation}
\end{widetext}
obtained by lineariazation of the Eq. (\ref{eq:LL}) with $\partial_{t}U=0$
at the homogeneous steady state solution $U=U_{0}$.

In the case when the high-order dispersion and spectral filtering
are absent $\beta_{3}=\beta_{4}=\delta=0$, Eq. (\ref{eq:characteristic})
gives two pairs of complex conjugate eigenvalues: 
\begin{equation}
\mu_{1,2}^{(0)}=\pm\sqrt{\theta-2I_{0}+i\sqrt{1-I_{0}^{2}}}\label{eq:lambda0}
\end{equation}
and $\mu_{1,2}^{(0)*}$, which determine the decay and oscillation
rates of the soliton tails. For example, for $S=2.0$ and $\theta=3.5$
we have $\mu_{1,2}^{(0)}=\pm\left(1.6837+0.275817i\right)$, which
means that in the absence of high-order dispersions the soliton tail
oscillations are strongly damped. This might explain the fact that without
soliton Cherenkov radiation it is hardly possible to observe soliton
bound state formation experimentally \citep{Leo10}. 

\begin{figure}
\includegraphics[scale=0.35]{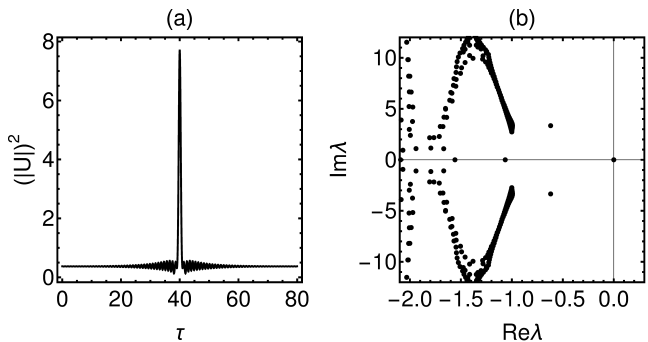}
\caption{Soliton solution of the Lugiato-Lefever equation (\ref{eq:LL}) with
$\beta_{4}=0.025$ and $\delta=0.02$ (a); eigenvalue spectrum obtained
by numerical linear stability analysis of this solution (b). Other
parameters are the same as in Fig. \ref{Fig0}.}
\label{Fig:soliton4}
\end{figure}

For nonzero but sufficiently small fourth-order dispersion coefficient,
$0<\beta_{4}\ll1$, the eigenvalues (\ref{eq:lambda0}) of Eq. (\ref{eq:characteristic})
are only slightly perturbed. However, in addition to (\ref{eq:lambda0})
two more pairs of complex conjugate eigenvalues, $\mu_{3,4}$
and $\mu_{3,4}^{*}$, appear. For zero spectral filtering coefficient, $\delta=0$, they are given
by 
\begin{equation}
\mu_{3,4}=\mp i\sqrt{\frac{1}{2\beta_{4}}\left[1+\sqrt{1-4\beta_{4}\left(2I_{0}-\theta+i\sqrt{1-I_{0}^{2}}\right)}\right].}\label{eq:lambdad=00003D00003D0}
\end{equation}
It is seen that real (imaginary) parts of $\mu_{3,4}$ in Eq.
(\ref{eq:lambdad=00003D00003D0}) vanish (diverge) in the limit $\beta_{4}\to0$.
When the spectral filtering coefficient is nonzero, $\delta>0$, analytical expressions for the eigenvalues $\mu_{3,4}$
become very cumbersome. However, in the limit $\beta_{4}={\cal O}(\delta)\ll1$
we get: 
\begin{eqnarray}
\mu_{3,4}&=&\mp\sqrt{\beta_{4}}\left[\frac{\sqrt{\left(1+\delta/\beta_{4}\right)^{2}-I_{0}^{2}}}{2}\right.\nonumber\\
&+&i\left.\left(\frac{1}{\beta_{4}}+\frac{\theta-2I_{0}}{2}\right)+{\cal O}\left(\delta\right)\right].\label{eq:lambda_asympt}
\end{eqnarray}
Due to the presence of the eigenvalues $\mu_{3,4}$ and $\mu_{3,4}^{*}$ the tails of the soliton of Eq. (\ref{eq:LL}) with $\beta_{3}=0$ and $0<\beta_{4}\ll1$ become weakly decaying and fast oscillating, which
favors the formation of soliton bound states, and can be referred to as the soliton Cherenkov radiation \citep{Akhmediev95}. Note, that when $\beta_{4}$ is sufficiently small, the term $\delta/\beta_{4}$
describing in Eq. (\ref{eq:lambda_asympt}) the contribution of spectral filtering into the real part of $\mu_{3,4}$ can lead to a considerable increase of the decay rate of the soliton tails without significant change of their oscillation frequency. For example, for $S=2.0$,
$\theta=3.5$, $\beta_{4}=0.025$ , and $\delta=0.02$ we get $\mu_{3}=-0.123-6.529i$, while for the same parameter set and $\delta=0$ one obtains $\mu_{3}=0.063-6.528i$. Numerically calculated intensity profile of the soliton solution of Eq. (\ref{eq:LL}) with small fourth-order dispersion coefficient $\beta_{4}=0.025$ is depicted in Fig. \ref{Fig:soliton4} together with the corresponding eigenvalue spectrum of the operator $\hat{L}\left(\mathbf{u}_{0}\right)$ defined by Eq. (\ref{eq:L}). 

Note, that the proof of the reflectional symmetry property of the discrete soliton
spectrum with respect to the $\Re\lambda=-1$ line given in \citep{barashenkov1996existence}
is trivially generalized to the case when even high order dispersions
are present. Nevertheless, the soliton spectrum shown in  Fig. \ref{Fig:soliton4}
does not possess this symmetry property due to the presence of nonzero spectral filtering coefficient $\delta =0.02$.  Furthermore,
as it is seen from Fig. \ref{Fig:soliton4}, for $\delta=0.02$
real parts of the complex conjugate eigenvalues, responsible for the
Andronov-Hopf bifurcation of the soliton, are shifted to the left from
the imaginary axis as compared to those shown in Fig. \ref{Fig:soliton0}
obtained for $\delta=0$. 

Sufficiently far away from the soliton core its trailing tail can
be represented in the form
\begin{equation}
u_{0}(\tau)\sim a_{3}e^{\mu_{3}\tau}+a_{4}e^{\mu_{3}^{*}\tau},  {\text{when}} \quad\tau\to+\infty,\label{eq:asa4}
\end{equation}
where the Cherenkov radiation amplitude $a_{3}$ is exponentially
small in the limit $\beta_{4}\to0$ \citep{Akhmediev95,Karpman93}, $a_{4}=p_{a}a_{3}^{*}$,
and for $\beta_{4}={\cal O}(\delta)\ll1$ we get
\begin{equation}
p_{a}=i\frac{1-\sqrt{1-I_{0}^{2}}}{A_{0}^{*2}\left(\frac{\delta}{\beta_{4}\sqrt{1-I_{0}^{2}}}+1\right)}+{\cal O}\left(\delta\right),\label{eq:p}
\end{equation}
where $p_{a}$ is independent of $\beta_{4}$ at $\delta=0$. Numerically
for $S=2.0$, $\theta=3.5$, $\delta=0.02$, and $\beta_{4}=0.025$
we obtain $p_{a}\approx 0.0571+0.0833i$.

{\section{Interaction between dissipative solitons}}.
Two or more solitons will interact through their overlapping oscillatory tails when they are sufficiently close to one another. In what follows, we investigate the interaction between two dissipative solitons. We consider the limit of weak overlap when the solitons are  well separated from each other and derive the interaction equations describing the slow time evolution of the soliton coordinates denoted by $\tau_{1,2}$. To this end, let us first rewrite Eq. (\ref{eq:LL}) in
a general form: 
\begin{equation}
\partial_{t}\mathbf{U}=\hat{F}\mathbf{U},\label{eq:general}
\end{equation}
where $\mathbf{U}=\left(\begin{array}{c}
U\\
U^{*}
\end{array}\right)$, $\hat{F}\mathbf{U}=\left(\begin{array}{c}
\hat{f}U\\
\hat{f}^{*}U^{*}
\end{array}\right)$, and $\hat{f}$ is the differential operator defined by the RHS
of Eq. (\ref{eq:LL}). We look for the solution of Eq. (\ref{eq:general})
in the form 
\begin{equation}
U(\tau,t)=U_{0}+u_{1}+u_{2}+\Delta u(\tau,t).\label{eq:ansatz}
\end{equation}
Here $u_{1,2}=u_{0}\left(\tau-\tau_{1,2}\right)$ are two unperturbed
soliton solutions, with slowly evolving in time coordinates $\tau_{1,2}\left(\varepsilon t\right)$,
$\Delta u(\tau,t)={\cal O}(\varepsilon)$ is a small correction to
the superposition of two solitons, and small parameter $\varepsilon$
describes the weakness of the overlap of the two solitons. Substituting
Eq. (\ref{eq:ansatz}) into the model equation (\ref{eq:general})
and collecting first order terms in $\varepsilon$ we obtain the following
linear inhomogeneous equation for $\mathbf{\mathbf{\Delta}u}=\left(\begin{array}{c}
\Delta u\\
\Delta u^{*}
\end{array}\right)$: 
\begin{equation}
\hat{L}(\mathbf{u}_{1}+\mathbf{u}_{2})\mathbf{\Delta u}=-\partial_{x}\mathbf{u}_{1}\partial_{t}\tau_{1}-\mathbf{\partial_{\mathbf{\mathrm{\mathit{x}}}}u_{\mathsf{\mathrm{2}}}}\partial_{t}\tau_{2}-\hat{F}(\mathbf{u}_{1}+\mathbf{u}_{2}),\label{eq:linear}
\end{equation}
where the linear operator $\hat{L}\left(\mathbf{u}\right)$ is defined
by Eq. (\ref{eq:L}). Due to the transnational invariance of Eq. (\ref{eq:LL})
this linear operator evaluated at the soliton solution $\mathbf{u}_{0}$
has zero eigenvalue corresponding to the so-called transnational neutral (or Goldstone) mode $\mathbf{v}_{0}=\left(\begin{array}{c}
v_{0}\\
v_{0}^{*}
\end{array}\right)$ with $v_{0}=du_{0}/d\tau$, $\hat{L}\left(\mathbf{u}_{0}\right)\mathbf{v}_{0}=0$.
The adjoint linear operator $\hat{L}^{\dagger}\left(\mathbf{u}\right)$
obtained from $\hat{L}\left(\mathbf{u}\right)$ by transposition and
complex conjugation also has zero eigenvalue with the eigenfunction
$\mathbf{w}_{0}=\left(\begin{array}{c}
w_{0}\\
w_{0}^{*}
\end{array}\right)$, which is referred below as the ``adjoint neutral mode'', $\hat{L}^{\dagger}\left(\mathbf{u}_{0}\right)\mathbf{w}_{0}=0$.
Below we will assume that $\mathbf{w}_{0}$ satisfies the normalization
condition $\left\langle \mathbf{w}_{0}\cdot\mathbf{u}_{0}\right\rangle =\int_{-\infty}^{\infty}\left(\mathbf{w}_{0}\cdot\mathbf{u}_{0}\right)d\tau=2\int_{-\infty}^{\infty}\Re\left(w_{0}^{*}u_{0}\right)d\tau=1$. 
\begin{figure}
\includegraphics[scale=0.4]{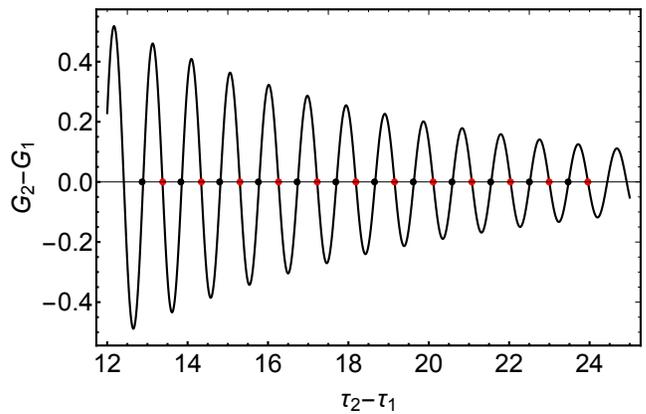}
\caption{RHS of Eq. (\ref{eq:intercat1}) as a function of the soliton separation $\tau_{2}-\tau_{1}$. Black (red) dots indicate the separations of the two solitons in stable (unstable) bound states calculated numerically.
Parameter values: $S=2.0$, $\theta=3.5$, $\delta=0.02$ and $\beta_{4}=0.025$.}
\label{Fig:interaction}
\end{figure}
Far away from the soliton core the asymptotic behavior of adjoint
neutral mode is given by: 
\begin{equation}
w_{0}(\tau)\sim b_{3}e^{\mu_{^{3}}^{*}\tau}+b_{4}e^{\mu_{3}\tau},\quad\tau\to+\infty,\label{eq:asw4}
\end{equation}
with $b_{4}=p_{b}b_{3}^{*}$, where asymptotic expression for $p_{b}$
coincides with that of $p_{a}$ given by Eq. (\ref{eq:p}).

\begin{figure}
\includegraphics[scale=0.34]{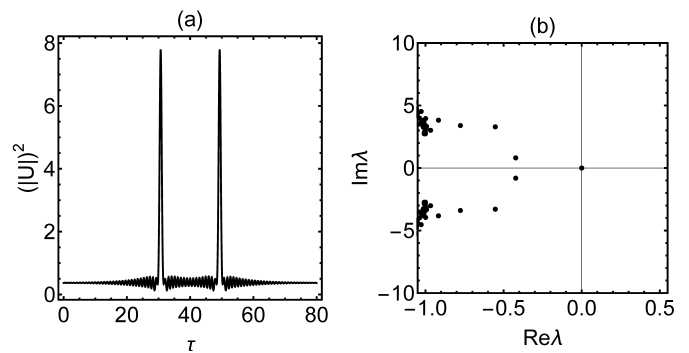}
\caption{Stable bound state of two dissipative solitons, $\delta=0.02$. (a) -- intensity distribution, (b) -- eigenvalue spectrum. Other parameter values are the same as for Fig. \ref{Fig:interaction}.} 
\label{Fig_BSs}
\end{figure}

\begin{figure}
\includegraphics[scale=0.33]{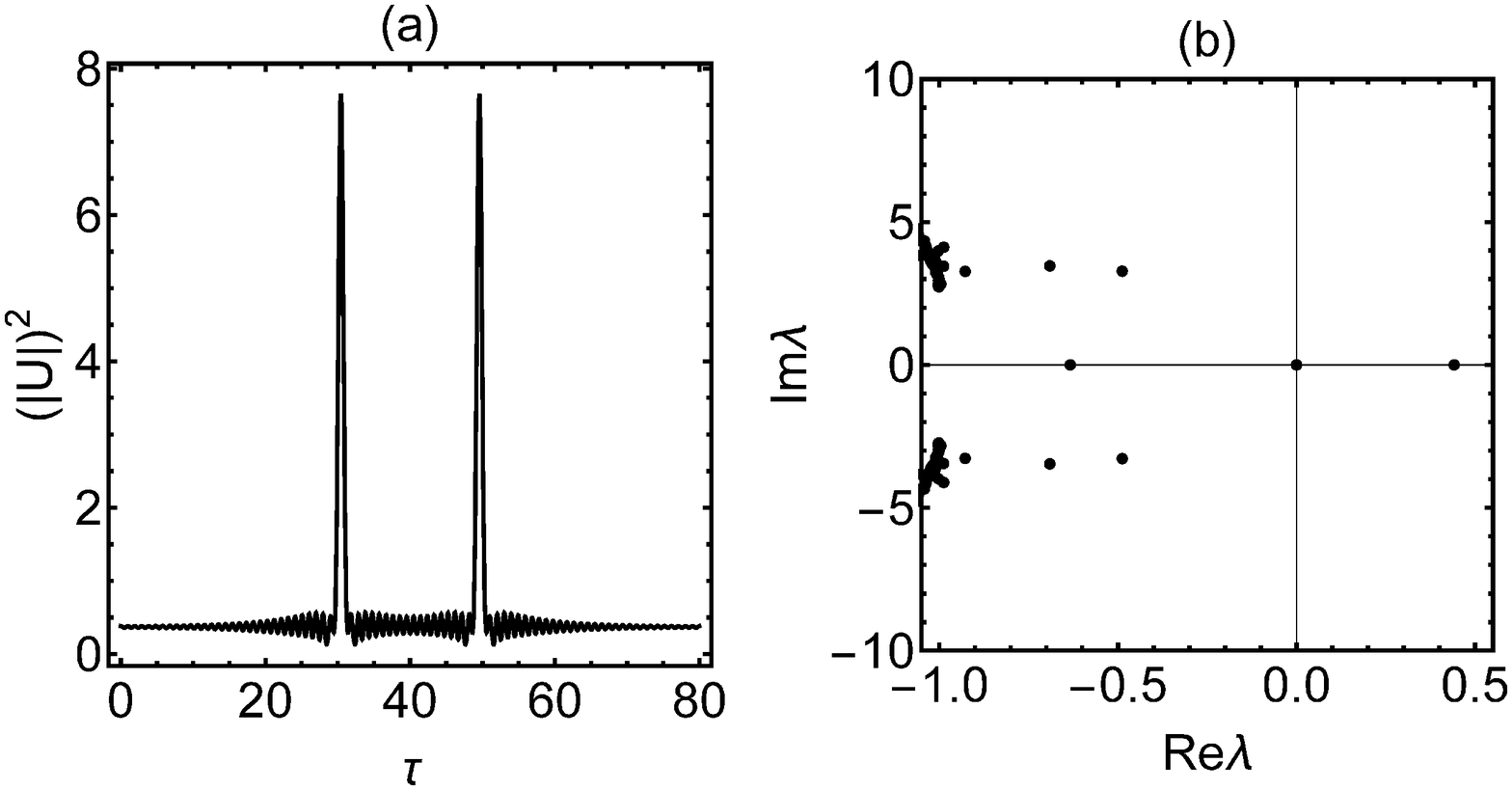}
\caption{Unstable bound state of two dissipative solitons, $\delta=0.02$. (a) -- intensity distribution, (b) -- eigenvalue spectrum. Other parameter values are the same as for Fig. \ref{Fig:interaction}.}. 
\label{Fig_BSu}
\end{figure}

\begin{figure}
\includegraphics[scale=0.34]{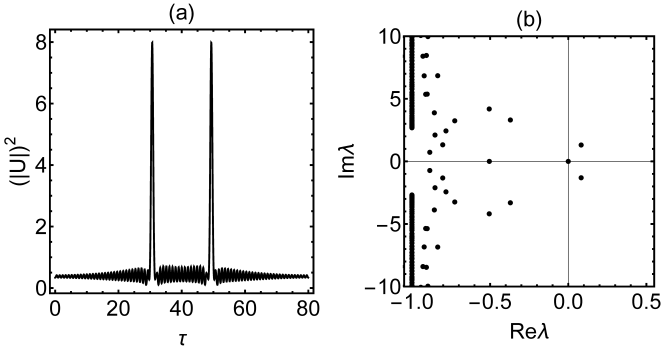}
\caption{The same bound bound state as shown in Fig. \ref{Fig_BSs}, but calculated for $\delta=0$. (a) -- intensity distribution, (b) -- eigenvalue spectrum. Bound state is unstable with respect to an Andronov-Hopf
bifurcation.} 
\label{Fig_BSS0}
\end{figure}

When the two interacting solitons are located sufficiently far away
from one another the solvability conditions of Eq. (\ref{eq:linear})
can be written as 
\begin{equation}
\partial_{t}\tau_{1,2}\approx G_{1,2},\quad G_{1,2}=\left\langle \mathbf{w}_{1,2}\cdot\hat{F}(\mathbf{u}_{1}+\mathbf{u}_{2})\right\rangle ,\label{eq:interact0}
\end{equation}
where we approximated the adjoint neutral modes of the operator $\hat{L}^{\dagger}(\mathbf{u}_{1}+\mathbf{u}_{2})$
by the adjoint neutral modes $\mathbf{w}_{1,2}=\mathbf{w}_{0}(\tau-\tau_{1,2})$
of the operators $\hat{L^{\dagger}}(\mathbf{u}_{1,2})$.

In order to derive the soliton interaction equations we need to calculate
$G_{1,2}$ in Eq. (\ref{eq:interact0}). To this end we split the
integral in Eq. (\ref{eq:interact0}) into two parts and using the
relations $\hat{L}^{\dagger}(\mathbf{u}_{1,2})\mathbf{w}_{1,2}=0$
together with the fact that $\mathbf{u}_{1}$ and $\mathbf{w}_{1}$
($\mathbf{u}_{2}$ and $\mathbf{w}_{2}$) are small for $\tau\in[0,+\infty)$
($\tau\in(-\infty,0]$), where the origin of coordinates $\tau=0$
corresponds to the central point between two solitons, $\left(\tau_{2}+\tau_{1}\right)/2=0$,
we get 
\begin{eqnarray}
G_{1,2}&=&\left\langle \mathbf{w}_{1,2}\cdot\hat{F}(\mathbf{u}_{1}+\mathbf{u}_{2})\right\rangle _{1,2}+\left\langle \mathbf{w}_{1,2}\cdot\hat{F}(\mathbf{u}_{1}+\mathbf{u}_{2})\right\rangle _{2,1} \nonumber \\
&\approx& \left\langle \mathbf{w}_{1,2}\cdot\hat{F}(\mathbf{u}_{1}+\mathbf{u}_{2})\right\rangle _{1,2}\nonumber\\
&\approx &\left\langle \mathbf{w}_{1,2}\cdot\hat{L}(\mathbf{u}_{1,2})\mathbf{u}_{2,1}\right\rangle _{1,2}-\left\langle \hat{L}^{\dagger}(\mathbf{u}_{1,2})\mathbf{w}_{1,2}\cdot\mathbf{u}_{2,1}\right\rangle _{1,2} \nonumber \\
&=&  \left(\delta+i\right)\left[\left\langle w_{1,2}\partial^{2}_{\tau}u_{2,1}\right\rangle _{1,2}-\left\langle u_{2,1}\partial^{4}_{\tau}w_{1,2}\right\rangle _{1,2}\right]\nonumber\\
&+&i\beta_{4}\left[\left\langle w_{1,2}\partial^{4}_{\tau}u_{2,1}\right\rangle _{1,2}-\left\langle u_{2,1}\partial^{4}_{\tau}w_{1,2}\right\rangle _{1,2}\right]+c.c.
\end{eqnarray}
with $\left\langle \mathbf{w}\cdot\mathbf{u}\right\rangle _{1}=\int_{-\infty}^{0}\left(\mathbf{w}\cdot\mathbf{u}\right)d\tau$, $\left\langle \mathbf{w}\cdot\mathbf{u}\right\rangle _{2}=\int_{0}^{\infty}\left(\mathbf{w}\cdot\mathbf{u}\right)d\tau$
and $\hat{L}^{\dagger}(\mathbf{u}_{1,2})\mathbf{w}_{1,2}=0$.

Next, performing integration by parts and using the symmetry properties
of the soliton and its neutral modes, $u_{0}(\tau)=u_{0}(-\tau)$,
$\partial_{\tau}u_{0}(\tau)=-\partial_{\tau}u_{0}(-\tau)$, $w_{0}(\tau)=-w_{0}(-\tau)$,
and $\partial_{\tau}w_{0}(\tau)=\partial_{\tau}w_{0}(\tau)$ we get:

\begin{eqnarray}
G_{1,2}&\approx \pm\left[\left(\delta+i\right)\left(w_{1,2}^{*}\partial_{\tau}u_{2,1}-u_{2,1}\partial_{\tau}w_{1,2}^{*}\right)\right.\nonumber\\
&+i\beta_{4}\left(w_{1,2}^{*}\partial^{3}_{\tau}u_{2,1}-u_{2,1}\partial^{3}_{\tau}w_{1,2}^{*}-\partial_{\tau}w_{1,2}^{*}\partial^{2}_{\tau}u_{2,1}\right.\nonumber\\
&+\left.\left.\partial^{2}_{\tau}w_{1,2}^{*}\partial_{\tau}u_{2,1}\right)\right]_{\tau=0}+c.c.=\nonumber \\
&\pm\left[\left(\delta+i\right)\partial_{\tau}(w_{0}^{*}u_{0})-i\beta_{4}\left(w_{0}^{*}\partial^{3}_{\tau}u_{0}+u_{0}\partial^{3}_{\tau}w_{0}^{*}\right.\right.\nonumber\\
&+\left.\left.\partial_{\tau}\left(\partial_{\tau}w_{0}^{*}\partial_{\tau}u_{0}\right)\right)\right]_{\tau=(\tau_{2}-\tau_{1})/2}+c.c.\label{eq:RHS}
\end{eqnarray}

Finally, substituting into Eq. (\ref{eq:RHS}) the asymptotic relations
(\ref{eq:asa4}) and (\ref{eq:asw4}) we obtain: 
\begin{widetext}
\begin{equation}
\frac{d\left(\tau_{2}-\tau_{1}\right)}{dt}\approx-\frac{12}{\sqrt{\beta_{4}}}e^{-\gamma(\tau_{2}-\tau_{1})}\Re\left[\left(1-i\frac{\delta}{3}\right)\left(a_{3}b_{3}^{*}e^{-i\Omega(\tau_{2}-\tau_{1})}-p_{a}p_{b}^{*}a_{3}^{*}b_{3}e^{i\Omega(\tau_{2}-\tau_{1})}\right)\right],\label{eq:intercat1}
\end{equation}
\begin{equation}
\frac{d\left(\tau_{2}+\tau_{1}\right)}{dt}=0,\label{eq:interact2}
\end{equation}
\end{widetext}
where $\gamma=\Re(\mu_{3})\approx\left(\sqrt{\beta_{4}}/2\right)\sqrt{\left(1+\delta/\beta_{4}\right)-I_{0}^{2}}$
, $\Omega=-\Im(\mu_{3})\approx1/\sqrt{\beta_{4}}+\sqrt{\beta_{4}}\left(\theta-2I_{0}\right)$,
and the Cherenkov radiation coefficients $a_{3}$ and $b_{3}$ are
exponentially small in the limit $\beta_{4}\to0$. For $S=2.0$, $\theta=3.5$,
$d=0.02$, and $\beta_{4}=0.025$ numerically we get $a_{3}\approx -0.158+0.149i$
and $b_{3}\approx 0.017+0.136i$. Finally, neglecting ${\cal O}\left(\delta\right)$
terms and taking into account that in the leading order in $\delta$ we have $p_{a}=p_{b}\equiv p$, Eq. (\ref{eq:intercat1}) can be rewritten
in the form: 
\begin{widetext}
\begin{equation}
\frac{d\left(\tau_{2}-\tau_{1}\right)}{dt}\approx\frac{12}{\sqrt{\beta_{4}}}e^{-\gamma(\tau_{2}-\tau_{1})}|a_{3}b_{3}|\left(|p|^{2}-1\right)\cos\left[\Omega\left(\tau_{2}-\tau_{1}\right)+\arg\left(b_{3}/a_{3}\right)\right].\label{eq:inter1}
\end{equation}
\end{widetext}
The RHS of Eq. (\ref{eq:inter1}) is plotted in Fig. \ref{Fig:interaction},
where the intersections of the black solid line with axis of abscissas correspond
to the soliton bound states. Examples of stable and unstable soliton bound states 
calculated numerically are shown in Figs. \ref{Fig_BSs}
and \ref{Fig_BSu}, respectively, together with the most unstable
eigenvalues of the operator $\hat{L}$ evaluated on the bound state
solutions.

Finally, in Fig. \ref{Fig_BSS0} we present the  same soliton bound state  as the one shown in Fig. \ref{Fig_BSs}, but calculated for $\delta=0$. It is seen that the eigenvalue spectrum of this state contains many discrete eigenvalues, which split from the continuous spectrum, and that it is oscillatory unstable due to the presence of two complex conjugate eigenvalues with positive real parts. Therefore, we can conclude that in the absence of spectral filtering the one-dimensional asymptotic equations (\ref{eq:intercat1})-(\ref{eq:inter1}) can be insufficient to describe the soliton interaction. 
The derivation of the interaction equations taking into account  an Andronov-Hopf bifurcation of the soliton bound states in the presence of fourth-order dispersion is beyond the scope of this study. A related problem concerning the effect of oscillatory instability on the soliton interaction was studied
in \citep{TVZ12}.

{\section{Conclusions}}.
We have considered an all fiber photonic crystal cavity coherently driven by an injected field. The intracavity field inside the fiber experiences self-phase modulation, dispersion, optical injection, and optical losses. Its space-time evolution can be described by the Lugiato-Lefever equation with high order dispersion, where, in addition, we have taken into account small spectral filtering term.  We have first discussed the properties of a single dissipative soliton and derived asymptotic expressions for the  soliton Cherenkov radiation amplitudes. We have focused our analysis on the regime, where the forth order dispersion and the spectral filtering coefficients are small, $0<\beta_{4},\delta\ll1$. Second, we have  investigated the interaction between two dissipative solitons in the case when they are well separated from each other. Assuming a weak overlap of soliton tails, we have established analytically the interaction law (Eqs. \ref{eq:inter1}) governing the slow time evolution of the coordinates of two interacting solitons. We have shown that although the Cherenkov radiation due to the small fourth-order dispersion can strongly enhance the soliton interaction and thus lead to the formation of a large number of soliton bound states, 
in the absence of spectral filtering these states can be unstable with respect to an oscillatory instability even when single soliton is well below the Andronov-Hopf bifurcation threshold. This means that one-dimensional equation (\ref{eq:inter1}) can be insufficient to describe the interaction of solitons in the generalized Lugiato-Lefever model (\ref{eq:LL}) with zero spectral filtering coefficient, $\delta=0$. On the other hand, the inclusion of small but sufficiently large spectral filtering, $0<\delta\ll1$, allows to stabilize oscillatory unstable bound states and validate the one-dimensional interaction equation (\ref{eq:inter1}).

\begin{acknowledgments}
We also acknowledge the support from the Deutsche Forschungsgemeinschaft (DFG-RSF project No. 445430311), French National Research Agency (LABEX CEMPI, Grant No. ANR-11- LABX-0007) as well as the French Ministry of Higher Education and Research, Hauts de France council and European Regional Development Fund (ERDF) through the Contrat de Projets Etat-Region (CPER Photonics forSociety P4S).  M. Tlidi is  a Research Director at the Fonds National de la Recherche Scientifique (Belgium). A. Vladimirov and M. Taki acknowledge the support from Invited Research Speaker Programme  of the Lille University.
\end{acknowledgments}
 
\bibliographystyle{apsrev}

\end{document}